\documentclass[10pt]{report}
\usepackage{array}
\usepackage{graphicx}
\begin{document}
{\bfseries \noindent Gravitational effects
on a rigid Casimir cavity}\\
\vspace{0.5cm} 
$E. \; Calloni^{1,2}$, $L. \; Di \; Fiore^{2}$, $G. \; Esposito^{2}$, 
$L. \; Milano^{1,2}$, $L. \; Rosa^{1,2}$  \\
\vspace{0.5cm}
\noindent 1: Universit\`{a} Federico II di Napoli, 
Complesso Universitario Monte Sant' Angelo, Via Cintia, 80126 Naples \\
2: INFN, Sezione di Napoli, Complesso Universitario Monte Sant'
Angelo, Via Cintia, 80126 Naples \\

\large \baselineskip=0.8cm

\font\bigbf = cmbx12

\font\bigbigbf = cmbx12 scaled \magstep2


\vspace{1.5cm}

\section*{{\bfseries Abstract}}
\vspace{0.5cm}
Vacuum fluctuations produce a force acting on a rigid 
Casimir cavity in a weak gravitational field. Such a force is here
evaluated and is found to have opposite direction with respect to the
gravitational acceleration; the order of magnitude
for a multi-layer cavity configuration is analyzed and
experimental detection is discussed, bearing in mind the current
technological resources.

\vspace{1.5cm}

In the last years a novel attention has been dedicated to
macroscopic evidence of vacuum fluctuations related to the Casimir
effect, both from a theoretical and experimental point of view,
with the aim to verify predictions and deviations from QED, or
to study deviations from Newtonian law at short distances
[1--4]. On the other hand, although a great theoretical effort
has been produced in order to understand vacuum fluctuations in a
gravitational field, laboratory tests are still completely 
lacking [5,6]. In this paper we compute 
the effect of a gravitational field on a ``rigid''
Casimir cavity, evaluating the net force acting on it, in order to
show that the order of magnitude of the resulting force, although
not allowing an immediate experimental verification, is
compatible with the current extremely sensitive force detectors,
actually the interferometric detectors of gravitational waves. The
Casimir cavity is rigid in that its shape and size remain unchanged under
certain particular external conditions such as the absence of accelerations
or impulses, constancy of temperature etc.,
and we evaluate the force by calculating the potential
of a cavity fixed in a Schwarzschild field, showing that it
explicitly depends on the gravitational potential. The associated
force turns out to have opposite direction with respect
to gravitational acceleration, and
orders of magnitude are discussed bearing in mind the current technological
resources as well as experimental problems.

In order to evaluate  the force due to the gravitational field
let us suppose that the cavity has geometrical configuration of two
parallel plates of proper area $A = L^{2}$ separated by the proper
length $a$. In Minkowski space-time the zero-point energy
of the system can be evaluated, in the case of perfect conductors, as
\begin{equation}
U = {\hbar}c L^{2}\sum_{n} \int \frac{d^{2}k}{(2\pi)^{2}}
\sqrt{k^{2} + \left({n\pi \over a}\right)^{2}},
\end{equation}
where we have introduced the normal modes
labelled by the positive integer $n$ and the transverse momentum $k$.
On performing the integral by dimensional regularization, the energy
takes the well known Casimir expression
\begin{equation}
U_{\rm reg} = -L^{2}\frac{\pi^{2}{\hbar}c}{720 a^{3}},
\end{equation}
where the final result is independent of the particular regularization
method. 

If the cavity is at rest in a static gravitational field, 
via the equivalence principle each normal mode
still resonates in the cavity, with associated energy reduced from
the gravitational red-shift by the factor $\sqrt{g_{00}}$; thus,
the total energy can be written as
\begin{equation}
U = {\hbar c}L^{2} \sum_{n} \int \frac{d^{2}k}{(2\pi)^{2}}
\sqrt{k^{2} + \left(\frac{n\pi \sqrt{g_{00}}}{a}\right)^{2}}.
\end{equation}
Evaluating again the integral by dimensional regularization, and
requiring that $g_{00} =  1 - \frac{\alpha}{r}$, 
with $\alpha \equiv {2GM\over c^{2}}$
as in a Schwarzschild geometry, we obtain (here we assume $L<<r$)
\begin{equation}
U_{\rm reg} = -L^{2}\frac{\pi^{2}{\hbar}c}{720 a^{3}}
\left(1 - \frac{\alpha}{r}\right)^{3/2}.
\end{equation}
The energy associated to the system turns out to be explicitly
dependent on the position in the gravitational field and is minimal
as $r \rightarrow \infty $. 

Now minus the gradient of $U_{\rm reg}$ with respect to $r$ yields
the force on the rigid cavity; for $r>> \alpha$, the resulting
expression can be approximated by
\begin{equation}
{\vec F} \approx \frac{\pi^{2} L^{2} \hbar c }{240a^{3}} 
\frac{g}{c^{2}} {\vec e}_{r},
\end{equation}
${\vec e}_{r}$ being the unit vector in the $r$-direction,
and $g$ the modulus of gravitational acceleration. Such a
force has opposite direction with respect to the
gravitational acceleration ${\vec g}$. The previous
expression can be rewritten as
\begin{equation}
{\vec F} \approx \frac{\pi^{2} L^{2} \hbar c }{240a^{4}} \frac{a
g}{c^{2}}{\vec e}_{r} = {\vec F}_{C}\bigtriangleup \phi, 
\end{equation}
where $\phi(r)$ is the gravitational potential. Thus, the force
exerted by vacuum fluctuations in a weak gravitational static
field can be interpreted as the ``difference in Casimir force'' on
the two flat plates. 

In considering the possibility of experimental verification of the
extremely small forces linked to this effect we begin by studying
the sensitivity reached by the present technology in detection of
very small forces on a macroscopic body, on earth, paying
particular attention to detectors of the extremely small
forces induced by a gravitational wave. In this case, as an
example, gravitational wave signals $h$ of order $h \approx
10^{-25}$, corresponding to forces of magnitude $ \approx
5 \cdot 10^{-17} N$ at frequency of few tens of Hz, are expected to be
detected with Virgo long baseline interferometric gravitational
wave detector presently under construction [7,8].

In the course of considering experimental possibilities of verification of the
force on a rigid Casimir cavity, we  evaluate this force on a
macroscopic body, having essentially the  same dimensions of
mirrors for gravitational wave detection and obtained through a
multi-layer sedimentation by a series of rigid cavities.
Each rigid cavity consists of two thin metallic disks, of
thickness of order 100 nm [9] separated by a dielectric
material, inserted to maintain the cavity sufficiently rigid:
introduction of a dielectric is equivalent to enlarging the optical
path length by the refractive index $n$ $(a \mapsto na)$. 
By virtue of presently low costs and facility
of sedimentation, and low absorption in a wide range of frequencies
[10], $SiO_{2}$ can represent an efficient dielectric material.

{}From an experimental point of view we point out that the Casimir
force has so far been tested  down to a distance $a$ of about
$60$ nm, corresponding
to a frequency $\nu_{min}$ of the fundamental mode equal to 
$2.5 \cdot 10^{15} Hz$. This limit is linked to the difficulty to
control the distance between two separate bodies, as in the
case of measurements of the Casimir pressure. As stated before, in
our rigid case, present technologies allow for cavities with much
thinner separations between the metallic plates, of the order of
few nanometers. At distances of order 10 nm, finite
conductivity  and dielectric absorption are expected to play an
important role in decreasing the effective Casimir pressure, with
respect to the case of perfect mirrors [11,12]. 
In this paper we discuss experimental
problems by relying  on present  technological resources,
considering cavities with plates' separation of 5 nanometers and
estimating the effect of finite conductivity by considering the
numerical results of Ref. [11]; for a separation of 6.5 nm this
corresponds to a decreasing factor $ \eta $  of about $7 \cdot 10^{-2}$
for Al. Last, to increase the total force and obtain
macroscopic dimensions, $N_{l} = 10^{6}$ layers can be used,
each having a diameter of 10 cm, and thickness of 100 nm, for a
total thickness of about 10 cm.

With these figures, the total force ${\vec F}^{T}$ acting on the body
can be calculated with the help of equation (6), modified to take
into account for $SiO_{2}$ the refractive index $n$, the decreasing
factor $\eta$, the area A of disk-shape plates, and the
$N_{l}$ layers:
\begin{equation}
{\vec F}^{T} \approx \eta N_{l} \frac{A \pi^{2} \hbar c }{240(na)^{3}}
\frac{ g}{c^{2}}{\vec e}_{r} , 
\end{equation}
whose magnitude is of order $10^{-15}N$.
Even though such a force is apparently more than one order of magnitude
larger than the force which the Virgo
gravitational antenna is expected to detect for a typical signal from
pulsars, we should bear in mind that the signal there is at reasonable high
frequency (some tens of Hz), while our calculated signal is
static. Indeed, the most important (and unsolved) experimental
problem in order to detect the force is the signal modulation [13-15],
and various experimental possibilities are currently under study. 

To sum up, we have found that the
effect of the gravitational field on a rigid  Casimir cavity
gives rise to a net force, whose direction is opposite to the one of
gravitational acceleration. Experimental verification of the calculated
force depends crucially on the capability of solving the signal 
modulation problem. On the other hand, in the authors'
opinion, the absolute value of the calculated force can be already
of interest to demonstrate that experiments involving effects of
gravitation and vacuum fluctuations are not far from the reach of current
technological resources.

\section*{{\bf Acknowledgments}}
We are indebted to Giuseppe Marmo for discussions
and to Serge Reynaud for correspondence. The work of
G. Esposito has been partially supported by the Progetto di Ricerca
di Interesse Nazionale {\it Sintesi 2000}. The INFN financial 
support is also gratefully acknowledged.

\section*{{\bfseries References}}

\noindent [1] M. Bordag, B. Geyer, G. L. Klimchitskaya and V. M.
Mostepanenko, {\it Phys. Rev.}  \textbf{D60}, 055004 (1999);
V. M. Mostepanenko and M. Novello, {\it Possible violation
of Newtonian gravitational law at small distances and constraints
on it from the Casimir effect} (HEP-PH 0008035).

\noindent [2] M. T. Jaekel and S. Reynaud, {\it Jour. Phys.}
{\bf I3}, 1093 (1993).

\noindent [3] K. Milton, {\it Dimensional and Dynamical Aspects of the
Casimir Effect: Understanding the Reality and Significance of Vacuum
Energy} (HEP-TH 0009173); G. Barton, {\it J. Phys.} {\bf A34},
4083 (2001).

\noindent [4] S. K. Lamoreaux, {\it Phys. Rev. Lett.} \textbf{83}, 3340
(1999); U. Mohideen and A. Roy, {\it Phys. Rev. Lett.} \textbf{81},
4549 (1998).

\noindent [5] S. A. Fulling, {\it Aspects of Quantum Field Theory in
Curved Space-Time} (Cambridge University Press, Cambridge, 1989).

\noindent [6] B. S. DeWitt, {\it Phys. Rep.} {\bf 19}, 295 (1975); 
M. Bordag, U. Mohideen and V. M. Mostepanenko, {\it Phys. Rep.}
{\bf 353}, 1 (2001).

\noindent [7] D. Blair, {\it The Detection of Gravitational Waves}
(Cambridge University Press, Cambridge, 1991).

\noindent [8] VIRGO Collaboration, {\it Nucl. Instr. Meth.} 
{\bf A360}, 258 (1995);
{\it The VIRGO Experiment} by VIRGO Collaboration, in:
Verbier 2000, Cosmology and Particle Physics, pp. 138--145.

\noindent [9] M. Bordag, B. Geyer, G. L. Klimchitskaya and 
V. M. Mostepanenko, {\it Phys. Rev.} {\bf D62}, 011701 (2000).

\noindent [10] R. A. Street, {\it Hydrogenated Amorphous Sylicon}
(Cambridge University Press, Cambridge, 1991); 
B. Chapman, {\it Glow Discharge Processes} (Wiley, New York, 1980).

\noindent [11] A. Lambrecht and S. Reynaud, {\it Eur. Phys. J.} 
{\bf D8}, 309 (2000).

\noindent [12] G. Jordan Maclay, {\it Phys. Rev.} \textbf{A61},
052110 (2000).

\noindent [13] S. Hunklinger, H. Geisselmann and W. Arnold,
{\it Rev. Sci. Instrum.} {\bf 43}, 584 (1972).

\noindent [14] P. R. Saulson, {\it Interferometric Gravitational Wave
Detectors} (World Scientific, Singapore, 1994).

\noindent [15] R. Esquivel--Sirvent et al., {\it Phys. Rev.} {\bf A}
(to appear in the November 2001 issue).

\end{document}